\documentclass{iau}
\usepackage{graphicx}

\title[Activity on a Li-rich giant: DI\,Psc revisited] 
{Activity on a Li-rich giant: DI\,Psc revisited}

\author[Levente Kriskovics et al.]   
{Levente Kriskovics,
Zsolt K\H{o}v\'ari,
Kriszti\'an Vida,
\and Katalin Ol\'ah}

\affiliation{Konkoly Observatory, \\
Konkoly Thege \'ut 15-17., H-1121, Budapest, Hungary \\ email: {\tt kriskovics}, {\tt kovari}, {\tt vidakris}, {\tt olah@konkoly.hu} \\[\affilskip]
}

\pubyear{2013}
\volume{302}  
\pagerange{100--101}
\jname{Magnetic Fields Throughout the Stellar Evolution}
\editors{A.C. Editor, B.D. Editor \& C.E. Editor, eds.}
\begin{document}

\maketitle

\begin{abstract}
We present a new Doppler imaging study for the Li-rich single K-giant DI\,Psc.
Surface temperature maps are reconstructed
for two subsequent rotation cycles. From the time evolution of
the spot distribution antisolar-type differential rotation pattern is revealed.
We show marks of non-uniform Li-abundance as well.
The possible connection between the current evolutionary phase of the star and its
magnetic activity is briefly discussed.

\keywords{stars: activity,
    stars: imaging,
    stars: individual (DI\,Psc),
    stars: spots,
    stars: late-type}
\end{abstract}

\firstsection 
\section{Introduction}
DI\,Psc (HD\,217352) is a rapidly rotating ($P_{\rm rot}=18.07$ days)
single K-giant, a new candidate for the small group of Li-rich K-giant stars. The extreme Li-abundance is related
to a short evolutionary episode, the helium flash, when different (partly unknown) processes activate
Li-production and propagation  \cite[(Charbonnel \& Balachandran 2000)]{charbonneletal2000}.
Moreover, the role of rotation and magnetic activity in these processes is also unclear.
Surface Li-abundance and the position of DI\,Psc on the HRD was determined in our recent
Doppler imaging study (\cite[K\H{o}v\'ari et al. 2013]{kovarietal2013}).
In this paper we aim to investigate the time evolution of the surface by Doppler imaging, as well as the Li-distribution on the surface.

\begin{figure}[b]
\begin{center}
 \includegraphics[width=4.5in]{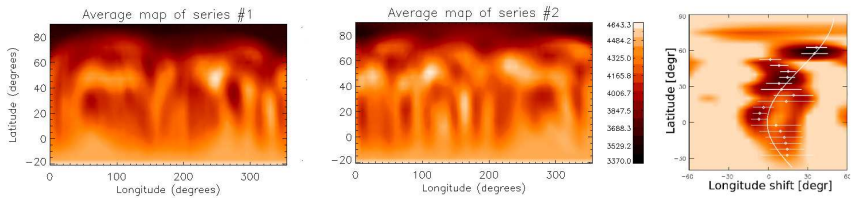} 
 \caption{Combined Doppler images for two consecutive rotation periods (left) in Nov-Dec 2012 and the fitted antisolar-type differential rotation on the cross-correlation function map (right).}
   \label{fig1}
\end{center}
\end{figure}

\section{Doppler imaging results}
15 time series spectra of exceptionally high signal-to-noise ratio were taken with \\* NARVAL@TBL in Nov-Dec 2012,
covering 40 days (i.e., $\approx2\times P_{\rm rot}$). We use the Doppler imaging code
\texttt{TempMap} by \cite[Rice et al. (1989)]{Ricetal89} for three lines (Fe\,{\sc i}-6430, Ca\,{\sc i}-6439 and Li\,{\sc i}-6708)
to reconstruct the stellar surface in two consecutive rotational cycles. Combined (Fe+Ca+Li) images,
as well as their cross-correlation is plotted in Fig.\,\ref{fig1}. The best fit correlation pattern suggests antisolar-type
surface differential rotation with a shear of $\alpha=-0.11\pm0.02$.
This result should be regarded as a preliminary one, since unexpected rearrangements in spot configuration (e.g., emerging new flux) can disturb seriously the pure differential rotation pattern, thus yielding false observation.
We note, however, that antisolar-type differential rotation was reported also for DP\,CVn, which is another Li-rich K-giant, the twin of DI\,Psc  \cite[(K\H{o}v\'ari et al. 2013)]{kovarietal2013}. High Li-abundance might be related to strong meridional circulation (i.e., extra mixing), which, on the other hand, can eventuate antisolar-type differential rotation (cf. \cite[Kitchatinov \& R\"udiger 2004]{kitchatinov2004}).

\begin{figure}[!!]
\begin{center}
 \includegraphics[width=4.5in]{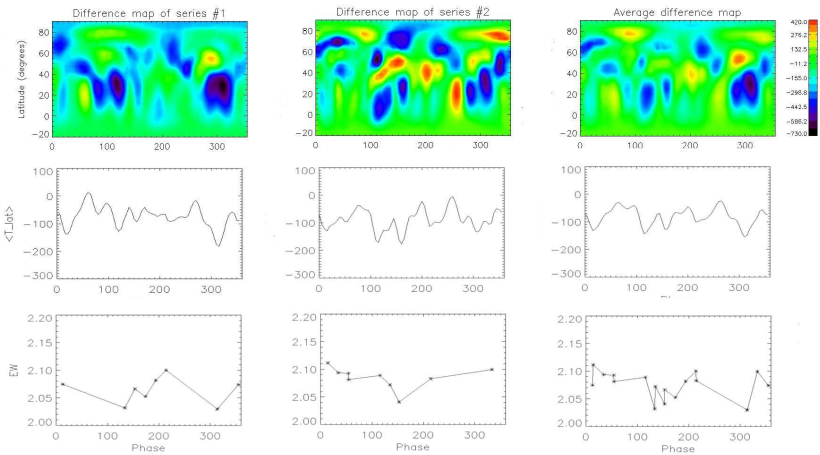} 
 \caption{Top: difference maps (i.e., Li-maps are subtracted from the average maps shown in Fig\,\ref{fig1} left). Note the similar structure for both rotational cycles. Middle: longitudinal distribution of the mean latitudinal temperature. Bottom: rotational modulation of the Li\,{\sc i}-6708
equivalent width.}
   \label{fig2}
\end{center}
\end{figure}

Non-uniform surface Li-abundance would affect the Doppler maps by altering the strength of the local line profiles (the temperature inversion code assumes constant abundance, therefore a higher Li equivalent width is fitted with a lower temperature causing a false cool spot on the map). In order to investigate this behaviour we subtract the Li-maps from the combined (Fe+Ca+Li) ones. The resulting difference maps are plotted in Fig.\,\ref{fig2} top, where signs of similar non-uniformity can be marked in both rotation cycles. This is supported by the longitudinal distribution of the mean latitudinal temperature values, as well as by the rotational modulation of the  Li\,{\sc i}-6708 equivalent width (middle and bottom panels of Fig.\,\ref{fig2}, respectively).
We conclude that DI\,Psc is an optimal target for studying the connection between surface activity, rotation and surface Li-enrichment.

\begin{acknowledgments}
This work has been supported by the Hungarian Science Research Program OTKA K-81421,
the Lend\"ulet-2009 and Lend\"ulet-2012 Young Researchers' Programs of the Hungarian Academy
of Sciences and by the HUMAN MB08C 81013 grant of the MAG Zrt.
\end{acknowledgments}

\end{document}